\documentclass[aps,prl,showpacs]{revtex4}
\usepackage{epsfig}
\usepackage[dvips]{color}     
\usepackage[latin1]{inputenc} 
\newcommand{\beq}{\begin{equation}}
\newcommand{\eeq}{\end{equation}}
\newcommand{\beqa}{\begin{eqnarray}}
\newcommand{\eeqa}{\end{eqnarray}}
\newcommand{\beqar}{\begin{eqnarray*}}
\newcommand{\eeqar}{\end{eqnarray*}}

\begin{document}
\thispagestyle{empty} \textheight 22cm
\title{Neutron-Proton Collisions }

\author{E. Di Grezia}%
\affiliation{\mbox{INFN, Sezione di Napoli,
Complesso Universitario di Monte S. Angelo}\\
Via Cintia, Edificio 6, 80126 Napoli, Italy}
\email{digrezia@na.infn.it}

\vspace{1cm}
\begin{abstract}

A theoretical model describing neutron-proton scattering developed
by Majorana as early as in 1932, is discussed in detail with the
experiments that motivated it. Majorana using collisions' theory,
obtained the explicit expression of solutions of wave equation of
the neutron-proton system. In this work two different models, the
unpublished one of Majorana and the contemporary work of Massey,
are studied and compared.
\end{abstract}

\maketitle


\section{Introduction}

In early 1932 a set of experimental phenomena revealed  that the
neutron plays an important role in the structure of nucleus like
the proton, electron and $ \alpha $-particle and can be emitted by
artificial disintegration of lighter elements. The discovery of
the neutron is one of the important milestones for the advancement
of contemporary physics. Its existence as a neutral particle
has been suggested for the first time by Rutherford 
in 1920 \cite{rut}, because he thought it was necessary to explain
the formation of nuclei of heavy elements. This idea was supported
by other scientists \cite{sci} that sought to verify
experimentally its existence. Because of its neutrality it was
difficult to detect the neutron and then to demonstrate its
existence, hence for many years the research stopped, and
eventually, in between 1928-1930, the physics community started
talking again about the neutron \cite{ros}.
For instance 
in \cite{ros} a model was developed in which the neutron was
regarded as a particle composed of a combination of proton and
electron. At the beginning of 1930 there were experiments on
induced radioactivity, which were interpreted as due to neutrons.
Indeed in 1930 Bothe and Becker \cite{bo} found that light nuclei
bombarded by $ \alpha $ particles produced radiation having higher
penetrating power than $\gamma$ radiation. In 1932 I. Curie and F.
Juliot \cite{curie} discovered the first artificial radiative
substances bombarding light elements, Beryllium and Boron, by
$\alpha$ particle (doubly ionized helium nuclei obtained from
spontaneous disintegration of polonium).

In the same period Chadwick made experiments \cite{ch}
 on the radioactivity and the experimental results obtained in these experiments and in those
 of I. Curie and F. Juliot, Bothe and Becker
 were explained by assuming that the radiation consisted of a new type of
particle of mass nearly equal to that of a proton and with no net
charge, i.e. the neutron.

The discovery of the neutron, raised a number of problems to be
analyzed:

1) the relation of neutron with electrons and protons;

2) emission of $\gamma$ rays associated with the neutron;

3) laws of interaction between neutron and nuclei.

This motivated the experiments of 1932 on disintegration of nuclei
of light elements with fast neutrons, on the conversion of their
kinetic energy in emission of $ \gamma $ radiation, on
distribution of speed in neutron scattering. After discovery of
the neutron in different laboratories the study of neutron
interactions with matter, in particular proton and electron,
continued \cite{dee, fea}. In fact 
the interaction of neutrons with electrons \cite{dee}, the
collision of neutrons with nitrogen nuclei \cite{fea}, the
concentration of slow neutrons in the atmosphere \cite{moo} were
examined and, in Italy, Rasetti was experimenting on Beryllium
\cite{ra}.

In this paper we attempt to give a short summary review about some
of these experiments including contributions of Feather and Dee.
As a matter of fact in order to understand these experiments one
has to take into account the laws of collisions of neutrons with
the matter.

Massey \cite{ma} and Majorana \cite{vol}, separately, proposed two
different models about the disintegration mechanism by neutrons.

In the two models proposed 
by Massey and Majorana, the spins of the particles have not been
considered, while later on Majorana proved that this is
significant in these interactions \cite{majio}.

The paper is organized as follows. In Sect. 1 we consider the
experiments by Chadwick, Feather, Dee, in Sect. 2 we give a brief
review on collision theory. In Sect. 3 we outline the proposal of
models, made by Massey and Majorana separately, for a description
of the passage of neutrons through matter in terms of the
collision theory. Finally in Sect. 4 we draw our conclusions.

\section{Collisions of Neutrons with atomic Nucleus. Experiments on Neutrons and their passage in the matter}

The nature and properties of the neutron are of the interest
because, as for the proton, it is important to understand the
structure of matter.

In fact the discovery of the neutron by Chadwick \cite{ch} is
followed by different experiments involving neutrons as
projectiles, to analyze nuclear structure. In the same period
Feather used neutrons as projectiles and found that they could
disintegrate the nitrogen nucleus \cite{fea}, while Dee studied
its interaction with electrons $e^-$ \cite{dee}.

The discovery by Chadwick of the neutron with mass approximately
that of the proton motivates different experiments to determine
the nature of this particle, i.e., the nature of the field
surrounding the particle. These experiments have investigated the
properties of the neutron trough interactions of the neutron with
material particles such as proton and atomic electrons. It is
therefore of interest to analyze the experiments on scattering of
neutron with the proton and neutron, and the theoretical
calculations of the behavior of neutrons in this respect. Hence in
the next two sections we are going to examine the principal
experimental and theoretical contributions about the neutron
collisions.

In this section we are going to analyze the experiments developed
in the same period by Chadwick, Feather and Dee to examine the
properties of the neutron which is taken into account in models
that we describe.

\subsection{Experiments of Chadwick}

In 1930, the German physicists Bothe and Becker \cite{bo}
bombarded the light metal beryllium with    particles, and noticed
that a very penetrating radiation was emitted. This neutral
radiation was non-ionising, and they assumed it consisted of rays.
In 1932 Irène and Frédéric Joliot-Curie \cite{curie} investigated
this Bothe's penetrating radiation in France. They let this
neutral radiation, generated by polonium-beryllium sources
irradiated by alpha particles, hit a block of paraffin wax, and
found it caused the wax to emit high speed protons (3-9 cm/s).
Because of the high speed of these protons, the $\gamma$ rays
would have to be incredibly energetic to knock them from the wax.
They interpreted the  resulting atomic recoils as Compton effect.
At the same time Chadwick \cite{ch} reported the Joliot-Curie's
experiment to Rutherford, who did not believe that gamma rays
could account for the protons from the wax. He and Chadwick
proposed that the beryllium was emitting a stream of neutrons,
which have nearly the same mass as protons, and hence should knock
protons from a wax block fairly easily. James Chadwick repeated
this experiment. The alpha-particles from the radioactive source
hit the beryllium nuclei and transformed them into carbon nuclei,
leaving neutral radiation (one free neutron). Then he used this
radiation to bombard hydrogen and nitrogen in the wax and it could
knock a free proton. He concluded that this neutral radiation was
absolutely not a gamma-radiation. Because gamma-radiation had no
momentum to produce proton from atom, i.e. Compton Scattering by
Gamma Rays would violate conservation laws, it was reasonable that
this radiation from beryllium was a kind of neutral particle with
a mass similar to the proton. The particle mass was estimated by
combining information from paraffin and nitrogen recoils and
nuclear decay measurements.

The first (I) step was to obtain a stream of neutrons and the
second step (II) was to detect it and then analyze its properties
(III).

Because of its neutrality the neutron has a great penetrating
power and it could be detected indirectly by ionization
measurements of recoiling nuclei, i.e. by collision of neutrons in
passage through the matter with an atomic nucleus. Although the
collisions were so infrequent, their number in a beam could be
estimated from the frequency of the collisions and the angular
distribution of the struck protons and electrons. Hence in its
passage through the matter the neutron is deflected from its path
because of the internal field of the nucleus. The struck nucleus
recoils and acquires energy to produce ions which can be detected
by a ionization chamber connected to an amplifier and
oscillograph.

The probability of a collision between a neutron and an atomic
nucleus depends on the number of neutrons and on their velocity.

In fact some experiments with slower neutrons suggested that the
radius for the proton-neutron collisions increased as much as the
velocity of the neutron decreased.

I) The first step is {\em to obtain neutrons} realized in a
nuclear reaction by a new nucleus formed by bombardment of
polonium $\alpha$-particles captured by neutron source, all the
light elements up to aluminum (lithium,beryllium, boron, neon,
fluorine, sodium, aluminum, magnesium), with exceptions of helium,
nitrogen, carbon, oxygen. The nuclear process according to the
nuclear reaction
\begin{equation}
(Z,A) + \alpha\rightarrow  (Z+2,A+3) + n
\end{equation}
consists in the capture of the $\alpha$-particle into the atomic
nucleus with the formation of a new nucleus and the release of a
neutron. The elements from which Chadwick obtained neutrons are
lithium, beryllium, boron, fluorine, neon, sodium, magnesium and
aluminum. The elements of higher atomic number up to argon
produced neutron if $\alpha$-particles had sufficient energy.

Consequently, in most experiments, beryllium and boron have been
used as sources of neutrons and the dependence of emission of the
neutrons on the velocity of the bombarding $\alpha$-particles has
been analyzed.



II) The second step is {\em to detect the neutron}. Chadwick
examined the dependence of the neutron emission on the velocity of
the bombarding $\alpha$-particles. Moreover, neutrons can be
detected only in an indirect way, by the observation of the
recoiling atoms, and Chadwick found that the probability of a
collision between an emitted neutron and a nitrogen atom in the
chamber depends on the velocity of the neutron, with less energy
of the recoil atoms when the neutron is slower.

In particular, Chadwick observed that the neutrons emitted from
beryllium by polonium $\alpha$-particles of velocity $1.6\times
10^9 $ cm/sec consisted of at least two groups: the slower with
velocity $2.8\times 10^9 $ cm/sec (energy of $4.1\times 10^6 $ eV)
and the faster group with velocity grater than $4\times 10^9 $
cm/sec (energy $>$ $8\times 10^6 $ eV) according to the reaction
\begin{equation}
Be_4^9 + He_2^4\rightarrow  C_6^{12} + n_0^1,
\end{equation}
and corresponding to these two groups of $n$ there are two groups
of recoil atoms.

Furthermore, the experiments of Chadwick provided evidence for the
emission of neutrons of energy up to about $12\times 10^6 $ eV.
Moreover, he showed that the capture of an $\alpha$-particle by
the beryllium nucleus results in a complete breakdown of the
nucleus, with the emission of an $\alpha$-particle, a neutron $n$,
and a $\gamma$-radiation. He measured that the mass of the neutron
was about the same as that of the proton, lying in between
$1.0058$ and $1.0070$, observing the momenta transferred in
collisions of neutron with atomic nucleus. Anyway, for an accurate
estimate of the mass of the neutron Chadwick used the energy
relation in the disintegration in which the neutron is ejected
from an atomic nucleus. In fact, assuming the conservation of
energy and momentum in the disintegration of the nucleus of the
known mass, the neutron mass is given from the measurement of the
kinetic energy of the neutron liberated by $\alpha$-particles of
known energy.

III) The third step is {\em to analyze the nature and the
properties of the neutron}. In this respect Chadwick obtained the
confirmation that the mass of the neutron was very close to that
of the hydrogen atom. This could be consistent with a model of the
neutron structure in which the neutron is combination of a proton
and an electron with binding energy corresponding to $1\times 10^6
$ eV. This argument could be in favor of the complex nature
(model) of the neutron. Hence it was necessary to understand the
nature of the neutron and its properties, realizing experiments of
nucleus collisions with neutrons, and if there was observation of
the splitting of the neutron into a proton and an electron to have
the most direct proof of the complex nature of the neutron. In
1935 the missing observation of the above property, the arguments
of quantum mechanics and relativistic mechanics and statistics,
and the measurement of nuclear spins, provided experimental proof
that the neutrons didn't contain electrons. All this supported the
idea that a neutron could be an elementary particle and the
hydrogen represented the only possible combination of a proton and
an electron. Before of this time, in 1932, different theoretical
models of the neutron had been proposed and different experiments
on the nuclear collision with neutron were made to confirm the
right nature, i.e. model, of the neutron.

In this context one should insert the experimental contributions
of Feather and Dee, and the theoretical contributions of Massey
and Majorana, that we will analyze in the next subsections.



In the next sections we will analyze the experiments of Feather
and Dee, which analyze the neutron interaction with protons and
electrons.


\subsection{Experiments on collisions of neutrons of
Feather and Dee}
In this section we analyze some experiments
involving the neutron collisions with proton and electron by which
it was possible to understand the neutron properties and nature.
In these experiments the neutron excited in light elements under
$\alpha$-particle bombardment interacts with the matter by means
of the expansion chamber or Wilson chamber.

For example, as we have said in the previous section, Curie and
Joliot \cite{curie}, in an expansion chamber, observed recoil
tracks of proton and helium nuclei from paraffin by neutrons from
Beryllium. Contemporarily Rasetti \cite{ra} and Auger \cite{au}
observed tracks of protons produced in the same way. On the other
hand, Faether and Dee
reported observations of proton and electron tracks. 

The neutrons have some very interesting properties described in
the papers of Chadwick, Feather, and Dee \cite{ch,dee, fea}. The
penetrating power of neutrons shows that they have no electric
charge, and the experiments proved their loss of energy is due to
the collisions with atomic nuclei and more rarely with the
electrons.

Moreover, these experiments confirmed that the neutron is a
particle
with mass $M$ nearly equal to that of the hydrogen atom, 
and investigated the relation between the scattering of neutrons
and their velocity. In particular, it was experimentally proved
that the faster neutrons are more easily stopped than the slower
ones, because the faster neutrons make more inelastic collisions
with the nuclei.

{\em Feather}, for the first time, made experiments of
disintegration of fluorine, nitrogen, oxygen, carbon nuclei. In
these experiments he measured the ranges of the atoms recoiling
and he found that the inelastic collisions were less frequent than
the elastic ones. Feather employed an automatic expansion chamber,
indispensable for such investigations on neutron interactions,
filled with nitrogen and traversed by neutrons produced by
beryllium under $\alpha$-polonium particle bombardment with
energies distributed over a wide range. In these experiments
Feather investigated collisions between neutrons and nitrogen
nuclei by stereoscopic photography of the produced recoil and
disintegration tracks. The tracks observed by Feather were the
result of elastic and inelastic collisions between neutrons of
mass $1$ and nitrogen nuclei. Feather observed about 130 cases of
neutron-nitrogen nucleus interaction, which are 109 elastic recoil
tracks (elastic collisions) of nitrogen nucleus projected at an
angle $\theta$ with the direction of incident neutron of a
definite initial velocity and the other 32 are disintegration
tracks (inelastic collisions). The inelastic collisions result in
artificial disintegration of nuclei under neutron bombardment, and
Feather measured they are of two main types:

1) in 12 occasions the neutron is captured with $\alpha$-particle
emission;

2) in 20 occasions there is no captured neutron.




The {\em capture cases} explored have been
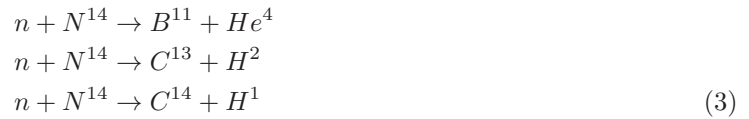
\begin{eqnarray}
&& n+ N^{14}\rightarrow B^{11} + He^4\nonumber \\
&& n + N^{14}\rightarrow C^{13} + H^2\nonumber \\
&& n + N^{14}\rightarrow C^{14} + H^1
\end{eqnarray}
the first one has been observed to take place in the forward and
reverse directions in about half the cases
\begin{equation}
 n+ N^{14}\leftrightarrow B^{11} + He^4,
\end{equation}
and $\gamma$-rays were emitted with the capturing of the neutron
and $\alpha$-particle emission. He found that the lengths of the
recoil tracks were in agreement with the neutron hypothesis of
Chadwick. Moreover, electronic tracks were also produced in the
Wilson chamber due to the passage of
neutrons. 




The {\em non-capture} cases explored have been
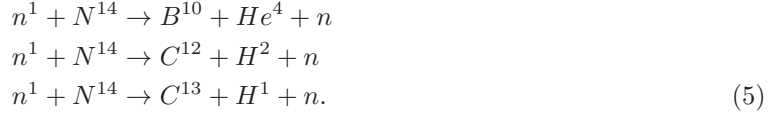
\begin{eqnarray}
&& n^1 + N^{14}\rightarrow B^{10} + He^4 +n\nonumber \\
&& n^1 + N^{14}\rightarrow C^{12} + H^2 +n\nonumber \\
&& n^1 + N^{14}\rightarrow C^{13} + H^1 +n.
\end{eqnarray}

Then in \cite{fea1} Feather investigated the action of neutrons
also on fluorine. He analyzed the disintegration of oxygen in a
Wilson chamber where neutron capture and $\alpha$-emission
occurred:
\begin{equation}
n^1 + O^{16}\rightarrow C^{13} + He^4 \end{equation}

Before the beginning of 1932 the only known nuclear disintegration
was that produced by 
$\alpha$-particles. Instead both $\alpha$'s and neutrons could
disintegrate the nucleus and light elements, in most cases,
captured by the incident particle, and ejected another one with
emission of $\gamma$-rays.

Thus, the experiments of Feather 
supported Chadwick in interpreting the penetrating radiation from
beryllium as consisting of neutrons.

 In this
artificial disintegration a general phenomenon is the emission of
$\gamma$-radiation. Moreover Feather undertook general
considerations of the energy changes in the various nuclear
processes and deduced the kinetic energy from measurements of
lengths of cloud tracks with range-velocity curves.

He also investigated range-velocity curves for the recoil atoms of
fluorine and carbon \cite{fea2}. In all these experiments there is
the evidence of recoil protons due to neutrons produced in the
resonance disintegration of beryllium and also of carbon recoil
atoms due to neutrons of high energy.


While Feather investigated, in an automatic expansion chamber, the
neutron collisions in its passage through matter with atomic
nuclei, producing recoil atoms of short range and great ionizing
power, on the other hand Dee \cite{dee} examined the interaction
of the neutrons, emitted by beryllium bombarded by
$\alpha$-particles of polonium, with electrons in a Wilson chamber
filled with Nitrogen atoms (gas of Nitrogen). The main conclusions
which Dee could draw from these experiments was that the
production of electron recoils by neutrons is a rare occurrence
compared with the production of recoil nitrogen atoms.

He obtained that the primary ionization along the path of a
neutron of velocity $3\times 10^9{\rm cm/sec}$ was of order of $1$
ion pair in $3$ meters of air, if the effective radius for a
neutron-nitrogen collision was $r=5\times 10^{-13}{\rm cm}$.
The main conclusion from these experiments was that the production
of electrons recoiled by neutrons was rare compared with proton or
nitrogen atoms recoil. Unlike the behavior of a proton which
dissipates its energy almost entirely in electron collisions, the
probability of interaction of a neutron with an electron is less
than $1\%$ of the probability of interaction with a proton or a
nitrogen nucleus.

These experiments show that the collision of a neutron with an
atomic nucleus is much more frequent than with an electron,
depending on the electric field between a neutron and a nucleus
small except at distances of the order of $10^{-12}$ cm.

Then they give information about the neutron radius. In fact if
the effective radius of neutron-nitrogen collision was of order
$5\times 10^{-13}$ cm then the ionization produced by a recoil
nitrogen atom along the path of neutron is less than 1 ion pair
per 3 meters of air with a neutron, ejected from a beryllium
nucleus, of velocity $3\times 10^9$ cm per second, while if the
radius was $5\times 10^{-12}$ cm the ionization can be  1 ion pair
per 3 cm of air. He observed the first case confirming that the
neutron radius is of order $5\times 10^{-13}$ cm.




\section{Neutron Models and Theory}
Taking into account the experimental results, the great
penetrating power of the neutron and its failure to interact with
the electron different models \cite{ma}, \cite{model}, \cite{ma1}
have been developed to report
on proton-neutron interaction. 


To understand the effect of the passage of the neutron through the
matter it is necessary to develop a theory with the help of wave
mechanics, assuming a possible potential neutron interaction
(neutron field) with other particles in accordance with the
experimental results. Hence at the beginning of 1932 many
theoretical physicists have tried to explain the existence of the
neutron and to establish a model that describes it.

One of the problems of quantum theory is to understand what is the
nature of the neutron, in order to explain the phenomena of
induced and natural radioactivity, artificial disintegration. In
1932 one option explored was to consider the neutron not as an
elementary particle, but as a combination of a proton and an
electron.

However there were conditions imposed by experiments:

a) the first hypotesis of Chadwick, following Rutherford
\cite{rut} that the neutron consisted of a proton and an electron;

b) the size of $ n $ is of order $ 10^{-12}-10^{-13} $ cm;

c) the mass of the neutron obtained from nuclear reaction $ B^{11}
+ \alpha \rightarrow N + N^{11}$, $m_n$ is in the $ \sim
1.005-1.007$

d) at the beginning of 1932 it was not yet clear whether or not
the neutron possessed. \footnote{ If one considered the neutron as
a complex particle there would be a problem with spin-statistics
Theorem. In 1935 the statistics and spins of lighter elements give
a consistent description if one assumes that the neutron is an
elementary particle \cite{ch1}, confirmed by spin-statistics
Theorem.}

On the basis of these conditions different models were proposed.
These models can be summarized as follows: (A) a dipole model of
strength $a\sim 10^{-13} e$ (like a dumbbell, with a positive and
negative charge separated by a small distance with their effects
cancelled), but this seems to be less probable than (B); (B) a
proton imbedded in an electron or an elastic spherically symmetric
model.

One of the models proposed represents the neutron as a dipole (A)
\cite{model} formed by two opposite charges at distance $ l $,
where the positive particle is the proton and the negative
particle is the electron with a potential:

\begin{equation} V = \frac{e \, l \cos{\theta}}{r^2}
\end{equation}
and a magnetic moment $m=l e$ that had to be measured
experimentally.

Another model \cite{model} (B)(a proton imbedded in an electron,
like an onion, with a sphere of one kind of electricity surrounded
by a layer of the other kind, so that again the charge is
cancelled) is a kind of reverse Thomson atom with positive charge
$ + $ at center in the negative charge density $ \rho (r) $
distribution. It has a spherical symmetry and represents well the
neutron suggested by Rutherford, and the potential at $ r $ is:

\begin{equation}
V=\frac{e}{r} - \frac{q(r)}{r},\,\,\,
\end{equation}
 where $q(r)=\int \rho(r)
dr$ or type

\begin{equation} V = f (r) e^{-kr} \end{equation} where $ f
(r)$ varies slowly with $ r $.

Hence in an elastic spherically symmetrical model the neutron is
like a hydrogen atom in a nearly zero quantum state, as was
discussed in 1920 by Rutherford \cite{rut}. This model was
discussed for the first time from the point of view of quantum
theory by Langer and Rosen \cite{ros1}. For such a system they
supposed the interaction between a material particle and a
possible neutron would be of the form $e^2 e^{-\lambda r}$, where
the parameter $\lambda$ is connected with the binding energy.

All these models should take into account the problem of
transition of neutron through the matter, obtaining different
theoretical results depending on the model.

Furthermore, one analyzes theoretically the phenomenon of passage
of the neutron through matter because the main effects of neutrons
are due to collisions of the neutron with atomic nuclei, more
rarely with electrons. As we have seen from the experimental point
of view \cite{dee, fea, fea1}, there are different types of
nuclear reactions involving neutrons (neutron interactions with
the matter) to reveal the neutron:

1) absorption of neutrons (calculation their absorption rate);

2) electron and proton scattering from a neutron (calculation of
the number of scattered neutrons distributed over the tracks);

3) disintegration by a neutron;

4) ionization from the neutron then calculating the loss of
ionization energy $ -dT/dx = f (T) $, which is a function of
kinetic energy T (calculation of the number of ions per unit path
and then the number of neutrons produced, depending on the model
of neutron and method of calculation).

Thus it must be a theory describing the phenomenon of neutron
collision with the matter, to explain the experimental results
\cite{fea, dee}. We must bear in mind that the theoretical results
relatively to experimental phenomena exhibit differences for
models of neutrons that are proposed. 

Massey \cite{ma} estimated $ f (T) $ in the case of spherical
model (B) where the neutron is considered as a particle composed
of a spherical symmetry with a mass about that of the proton and
with a potential field outside $ V (r) = \frac{e}{r} e^{-kr} $
applying the theory of Born collisions, which we briefly review in
the next section. He then obtained the effective cross-section and
evaluated the energy loss with Bethe's method.

There is another model proposed by Majorana in his unpublished
work in which the neutron is considered as an elementary particle.

We will analyze these two models in next sections, with a brief
review on the theory of collisions that is at the basis of the
models of Massey and Majorana.

Hence a study of the angular distribution of recoil tracks leads
to important data for a theory of the field of the neutron. In
fact the results of the experiments made to determine the field
force consisting in the observations of the collisions of neutrons
with material particles such as protons and electrons, have to be
interpreted. All this requires the development of a theory of such
collisions. The smallness of the field interaction between a
neutron and a charged particle leads to the possibility of
applying the approximate quantum theory of collisions of Born
\cite{born} in elastic scattering 
of neutron with particles, as Massey did.

In the next section we will review the basic properties of
collisions theory that Massey applied to elastic collisions of
neutrons with material particles. In fact we will highlight in the
work of Massey that, knowing the laws of collisions of neutrons
with matter, he could interpret the experimental data to determine
the field of a neutron in nuclear collisions, confirming that the
radius of such a particle is less than $2\times 10^{-13}$ cm in
the case of a particular model of neutron.






\section{Quantum Collision Theory}
To understand the experimental results on the collisions it is
necessary to use scattering theory. In this section we give a
brief summary about collision theory that is at the base of
neutron interactions models and of models proposed by Massey and
Majorana. Suppose that we have a stream of $N$ particles per unit
area per second incident with velocity $v$ on particle-target. The
number of particles deflected between angles $\theta$ and $\theta
+ {\rm d}\theta$ is $2\pi N {\rm I}({\rm\theta})\,{\rm
\sin}{{\rm\theta}}{\rm d}\theta$, and the collision cross-section
$Q$ is given in terms of ${\rm I}(\theta)$.

In fact collisional processes are described quantitatively in
terms of cross sections and to study them one needs the quantum
mechanics. One can distinguish between elastic and nonelastic
collisions depending on whether or not translational momentum and
energy are conserved. In this section we recall the main steps of
the quantum collision theory in the case of two elastic
interacting particles.As we said the main observable quantity
involved in a scattering process is the collision cross-section
$Q$ given by
\begin{equation}
Q= \int_0^{2\pi} \int_0^\pi I(\theta,\phi) {\rm \sin}{\theta}{\rm
d}\theta {\rm d}\phi= \int_0^{2\pi} \int_0^\pi |f(\theta,\phi)|^2
{\rm \sin}{\theta}{\rm d}\theta {\rm d}\phi
\end{equation}
where the function $I(\theta)$ has the dimension of area and, in a
scattering of a particle of mass $m$ and velocity $v$ by a
potential $V(r)$, is given by the exact formula
\begin{equation}
I(\theta)=|f(\theta,\phi)|^2=\frac{1}{k^2}\left|\sum_n(2n
+1)(e^{i\delta_n}-1)P_n ({\rm \cos}{\theta})\right|^2,
\label{exact}
\end{equation}
where $k=2\pi m v /h$ is the wave number and  $\delta_n$ are the
unknown phases, which are important for the purpose of evaluating
$Q$. Hence in the elastic collision theory one considers the wave
equation of Schrodinger for the relative motion of two interacting
particles of comparable masses $M_1$, $M_2$
\begin{equation}
\nabla \psi + \left\{ k^2 - \frac{8\pi^2 m}{h^2}V(r)
\right\}\psi=0, \label{a}
\end{equation}
where $M=\frac{M_1M_2}{M_1 + M_2}$ is the reduced mass of the
system of colliding particles. In the collision theory one
requires that a proper solution of the equation (\ref{a}) has the
asymptotic form \cite{cross}
\begin{equation}
\psi\sim e^{ikr\cos{\theta}} + f((\theta,\phi))\frac{e^{ikr}}{r}
\end{equation}
and to obtain such a solution $\psi$ is expanded
\begin{equation}
\psi=\sum_n \psi_n(r)P_n({\rm \cos}(\theta))
\end{equation}
where $\psi_n$ must then satisfy the wave equation subject to
finiteness at the origin
\begin{equation}
\frac{d^2 r\psi_n}{dr^2} + \left\{ k^2 - \frac{8\pi^2 m}{h^2}V(r)
-\frac{n(n+1)}{r^2}\right\}(r\psi_n)=0, \label{bb}
\end{equation}
while having the asymptotic form \cite{cross}
\begin{equation}
\psi_n\sim \frac{A_n}{k r}{\rm \sin}(k r - \frac{1}{2}n\pi
+\delta_n).\label{f}\end{equation} From (\ref{f}) Faxen and
Holtsmark \cite{cross} obtained the following expression for the
collision cross-section $Q$
\begin{equation}
Q=\frac{4\pi}{k^2}\sum_n(2n +1) {\rm \sin}^2\delta_n. \label{aa}
\end{equation}
Thus to obtain $Q$ it is necessary to calculate the phases
$\delta_n$. In collision theory there are two methods to calculate
the phases: the method of Jeffrey for $\delta_n$ greater than
unity, and the method of Born when $\delta_n$ is less than unity.

Hence if the scattering potential field $V(r)$ is small compared
with the centrifugal force term, i.e.
\begin{equation}
\frac{8\pi^2 m}{h^2}V(r) \ll \frac{n(n+1)}{r^2} \label{19}
\end{equation}
for $r$ such  that $kr\sim n + \frac{1}{2}$, for large $n$ and the
phase $\delta_n$ is small ($\delta_n\ll 1$) hence one is in the
validity regime of Born's approximation \footnote{There is a
Theorem for which a class of Lebesgue-summable on $R^3$ potential
V, the Rollnik $class_R$ potential \cite{roll} such that the Born
series is convergent.}, \cite{born} and the exponential in
(\ref{exact}) can be expanded in series and one obtains the first
approximation of Born method. In fact under these conditions Mott
\cite{mott} showed that $\delta_n$ has the following approximate
expression
\begin{equation}
\delta_n= \frac{4\pi^3 m}{h^2} \int_0^\infty V(r)
J_{n+\frac{1}{2}}^2(kr)r{\rm d}r \label{20}
\end{equation}
where $J_{n+\frac{1}{2}}$ are the Bessel functions of half-odd
order. Then assuming ${\rm \sin}\delta_n\sim  \delta_n$ (for
$\delta_n\ll 1$) it is possible to sum the series (\ref{aa}) for
$Q$ to give
\begin{equation}
Q=  \frac{64\pi^4 M^2}{h^4}\int_0^\pi
\left|\int_0^{\infty}V(r)\frac{{\rm \sin}(2kr {\rm
\sin}\frac{1}{2}\theta)}{2kr {\rm \sin}\frac{1}{2}\theta}r {\rm
d}r\right|^2{\rm \sin}{\theta}{\rm d}\theta \label{b}
\end{equation}
which is the expression of $Q$ in the approximation due to Born
\cite{born} (the formula of Born). In this approximation the total
collision cross-section $Q$ is finite if $V(r)$ vanishes at
infinity faster than $r^{-3}$, the same condition is for the exact
formula (\ref{exact}), because for $n$ sufficiently large the
exact and approximate series (\ref{exact}), (\ref{aa}) converge
together by virtue of (\ref{19}) and (\ref{20}). In the case in
which the Born approximation is not applicable ($\delta_n\gg 1$),
i.e. for small $n$, for
 $\frac{8\pi^2 m}{h^2}V(r) \gg
\frac{n(n+1)}{r^2}$, the phases $\delta_n$ can be calculated by
using an approximation based on classical theory given by Jeffreys
\cite{jeff} (method of Jeffreys or JWKB method). This method gives
for the solutions of the equation (\ref{bb}) the following
asymptotic forms
\begin{equation}
\psi_n\sim{\rm \sin}\left(\frac{\pi}{4}+ \int_{r_0}^\infty
f^{1/2}(r){\rm d}r\right),\,\,\, {\rm \sin}\left(\frac{\pi}{12}+
\int_{r_0}^\infty f^{1/2}(r){\rm d}r\right) \label{c}
\end{equation}
where $f(r)= k^2 - \frac{8\pi^2 m}{h^2}V(r) -\frac{n(n+1)}{r^2}$
and $r_0$ is the largest zero of $f(r)$, and only the first is the
required solution because it is finite at the origin. Comparing
(\ref{c}) with (\ref{f}) the phase $\delta_n$ is
\begin{equation}
\delta_n= \frac{1}{2}n\pi + \frac{1}{4}\pi +
\int_{r_0}^\infty[f^{1/2}(r)-k]{\rm d}r.
\end{equation}
For the case of intermediate phase, $\delta_n$ can be obtained by
interpolation of the two previous methods.

In conclusion, the conditions on $V(r)$ imply the use of one of
two methods, hence they provide a way to choose the mutual field
force
 $V(r)$.

 In a system which satisfies certain symmetry properties it is
possible to have some simplification in computation of the phases.
For example  Mizushima was the first to find the solution for the
scattered wave  \cite{miz} in an elastic sphere model (collisions
between rigid spheres) with the interaction energy $V(r)$
\begin{displaymath} V=\left\{\begin{array}{ll}
& \infty , \,\,\,for\,\,\, r< R, \\
& 0 , \,\,\,\,\,\,\,\,for \,\,\,r > R,
\end{array}\right.
\end{displaymath}
and it is given by
\begin{equation}
\psi = -i C \sum_{n=0}^{\infty} (2n +1)\frac{ J_{n+\frac{1}{2}}(k
R) }{ H^{(2)}_{n+\frac{1}{2}}(k R)} P_n (\cos{\theta})\frac{R
\,\,e^{-ik r}}{ r}
\end{equation}
where $k = 2\pi m V/h =2\pi \lambda$, $R$ is interpretable, in a
spherical model for interaction between neutron and nucleus, as
the nuclear radius plus neutron radius, and $J_{n+\frac{1}{2}}$ is
a Bessel function and $H^{(2)}_{n+\frac{1}{2}}$ is a Hankel
function of the second kind, while $P_n (\cos{\theta})$ is a
Legendre polynomial. In particular Massey and Mohr in the appendix
of their paper \cite{masseymohr} obtained the cross section $Q$ in
this model and using the general solution of the wave equation
(\ref{bb})
\begin{equation}
r^{-1/2}\psi= A J_{n+\frac{1}{2}}(kr) + B J_{-n-\frac{1}{2}}(kr)
\label{sol}
\end{equation}
 the phases are \begin{equation}\delta_n =(-1)^{n+1}{\rm
arctan}\frac{B}{A},\label{33}
\end{equation}
obtained from the condition that the solution (\ref{sol}) is zero
at $R$. Substituting the expression of $\delta_n$ (\ref{33}) in
(\ref{aa}) one obtains $Q$.


\subsection{Application of Collision Theory to neutron scattering in a model proposed by Massey}

In this section we outline the model formulated by Massey to
describe the neutron behavior in the collisions with the other
particles. Massey applied the quantum theory of collision due to
Born \cite{born}, Faxen and Holtsmark \cite{cross} to the
collision of neutrons with atomic nuclei and with electrons. He
considered the model of neutron in which the neutron is viewed
like an atom consisting of a proton and an electron (the spherical
{\em model (B)} in the previous section), suggesting a complex
nature of the neutron, i.e. a hydrogen atom in a zero quantum
state, in which the electron moves in a field given by
\begin{equation}
V(r) = e^2 \left(\frac{1}{r} +
\frac{Z}{a_0}\right)e^{-\frac{Zr}{a_0}} \label{mass}
\end{equation}
where $Z$ is the effective nuclear charge, $a_0$ is the Bohr
radius, and $a_0/Z$ is the neutron radius. 
Massey assumed that the electron and proton behave as point
charges and considered the collisions of neutrons with these
particles as the scattering of particles by a potential field
$V(r)$, i.e. the potential interaction field of a proton-neutron
can be compared to a deep hole of small radius. He showed that the
model assumed for the neutron was able to explain the experimental
results such as that its radius must be less than $2.0\times
10^{-13}$ cm and the probability of disintegration of a neutron by
nuclear collision is very small for this model.

In particular Massey, applying the collision theory, found that,
if the proton behaves as an elementary charge at small distances
of interaction, the collision cross-section is given by

\begin{equation}
Q\sim \frac{ 16\pi^5 M^2 e^4 a_0^4}{h^4 Z^4}
\end{equation}
where $M$ is the proton mass, and he found that the collision
radius in this model must be less than $1.4\times 10^{-14}$ cm,
against the experimental observations which indicate a greater
value than this as Feather measured \cite{fea}.

Hence this could induce to think that the proton could not behave
as a unit charge for distances of interaction less than $10^{-13}$
cm, either the energy of interaction between neutrons and nuclei
at large distances increases less rapidly, or the neutron could be
a point neutral particle.

In this context it is possible to insert the contribution of
Majorana, that we will analyze in the next section. To conclude
this section we outline the study of Massey about the
neutron-electron collision. He found that the number of ions
produced per centimeter of path by neutrons is very small, because
in its passage through air, containing $5.3\times 10^{20}$
electrons per $cm^3$, the total number of neutron-electron
collisions per centimeter path is of order $3.0\times 10^{-13}$
per cm, from which the total number of ions formed per centimeter
path could be less than $1$ ion per $10^{10}$ cm, explaining
therefore only the negative results of experiments of Dee about
the electron-neutron collision, but not the numerical
correspondence \cite{dee}.

The spherical model of Massey is valid as long as the Born formula
is valid and the wave length associated with the collision is long
compared with the size of the neutron, and the field of the
neutron must vary gradually.

Another model alternative to this one, that Massey investigated,
was a dipolar model ({\em model (A)} in the previous section) in
which the neutron could behave as a dipole with potential field
\begin{equation}
V(r) = \alpha \frac{e^2}{r^2}.
\end{equation}
It could be possible to distinguish the spherical or dipolar
distribution from the experimental point of view by measuring if
the interaction falls off more rapidly than $r^{-1}$ or $r^{-2}$
for large r.

Massey deduced that the spherical model was better suited for
describing some experimental results than the dipolar. The
''complex structure" of neutron proposed by Massey could be at the
base of some negative theoretical results as we said above.

In fact a proof that supported the validity  of the model of
complex nature of the neutron could be the observation of the
splitting of the neutron into an electron and a proton in a
nuclear scattering, but this wasn't observed in the experiments.
Moreover a difficulty of this model, if one considered the spin of
particles, was the inconsistency of statistics and spin
\cite{ch1}.


\subsection{
Model proposed by Majorana about passage of neutrons through
matter}

The model of neutron of a complex particle, proton-electron
combination, used by Massey in its calculations seems to be
 in disagreement with some experimental
observations (see for example \cite{dee},\cite{fea},
\cite{curie2}).

The direction in which either of {\em the models A, B} of the
neutron would eject protons was calculated and it was found that
the dumbbell type should eject them all perpendicularly to their
own path, while the onion type would eject some straight ahead,
with about ten times as many being thrown off perpendicularly.

Experiments with neutrons did not confirm either of these models
and hence the neutron is not built according to either of the
accepted models.

Thus, motivated by these results, one could regard the neutron as
an elementary particle rather than as a composite particle, as
like the electron and proton.

Indeed it is possible to formulate another model in which the
neutron is considered an elementary particle and there is another
type of
exchange interaction between neutron and proton. 

Thus, the innovative intuition of Majorana was to develop a first
theoretical model of neutron different from models (A), (B), in
which in the first instance the spin of proton and neutron was
neglected (a neutron model as an elementary particle of neutron
with spin was analyzed by Majorana in the published paper of 1933
\cite{majio}) and proton and neutron are considered as {\em
elementary particles}, in order to analyze the relative motion
between a neutron and a proton in a scattering process. In fact
this intuition will be confirmed from experimental results that
are explained if the neutron is not a mere close combination of
electron and proton acting like a fundamental particle of nature,
but it actually is an elementary particle itself. He studied the
following radial wave equation for relative motion of the neutron
and nucleus:
\begin{equation}
u'' + \frac{2}{r} u' + \left(\frac{2m}{\hbar^2}(E-V) -
\frac{l(l+1)}{r^2}\right)u=0 \label{1}
\end{equation}
where $m \sim 1/2 M_N$ is the reduced mass of the system.

He assumed that the field of interaction between a neutron and the
proton was a square potential well unlike the choice of Massey
(\ref{mass}):
\begin{displaymath}
V=\left\{\begin{array}{ll}
& -A  \,\,\,\,\,\,\,\,{\rm for}\,\,\,\,\,\,\,\, r< R, \\
& \,\,\,\,\,\,0  \,\,\,\,\,\,\,\,{\rm for} \,\,\,\,\,\,\,\,r > R.
\end{array}\right.
\end{displaymath}
For $r<R$, a solution of Eq. (\ref{1}), regular at the origin, was
\begin{equation}
 u = \frac{1}{\sqrt{r}} {\cal I}_{l+1/2}\left(\sqrt{\frac{2m}{\hbar^2}(E+V)}\,r\right), \;\;\; r< R,  \label{2}
\end{equation}
(${\cal I}_{l+1/2}$ are the modified Bessel functions of first
kind)while for $r>R$, the solutions were linear combinations of
half-odd order Bessel and Neumann functions
\begin{eqnarray}
&&\frac{1}{\sqrt{r}} {\cal I}_{l+1/2}\left(\sqrt{\frac{2m}{\hbar^2}E}\,r\right),\nonumber \\
&&\frac{1}{\sqrt{r}} {\cal
N}_{l+1/2}\left(\sqrt{\frac{2m}{\hbar^2}E}\,r\right),
\end{eqnarray}
with the boundary condition that the solution reduced to (\ref{2})
at $r=R$. He obtained the following solution of Eq. (\ref{1}),
which was regular at the origin:
\begin{displaymath}
u_l=\left\{\begin{array}{ll} & \frac{C_l}{\sqrt{r}} {\cal
I}_{l+1/2}(k_0 r)
\,\,\,\,\,\,\,\,\,\,\,\,\,\,\,\,\,\,\,\,\,\,\,\,\,\,\,\,\,\,\,\,\,\,\,\,\,\,\,\,\,\,\,\,\,\,\,\,\,\,\,\,\,\,r < R , \\
& \frac{C_l}{\sqrt{r}} \left(a{\cal I}_{l+1/2}(k r) +b{\cal
N}_{l+1/2}(k r)\right) \,\,\,\,\,\,\,r > R.
\end{array}\right.
\end{displaymath}
where
\begin{equation}
 k^2 = \frac{2m}{\hbar^2}E, \,\,\,\,\,\,\,\,\,\,\,\,\,\,\,\,\,\, k^2_0 = \frac{2m}{\hbar^2}(E+V), \label{3}
\end{equation}
and 
$a=a(x),b=b(x)$ were:
\begin{eqnarray}
&&a =\frac{\pi x}{2} \left({\cal I}_{l+1/2}(k_0r){\cal
N}'_{l+1/2}(k r)-\frac{k_0}{k}{\cal I}'_{l+1/2}(k_0 r){\cal
N}_{l+1/2}(k r)\right),
\nonumber \\
&&b =\frac{\pi x}{2} \left(\frac{k_0}{k}{\cal I}_{l+1/2}(k r){\cal
I}'_{l+1/2}(k_0r)-{\cal I}'_{l+1/2}(kr){\cal N}_{l+1/2}(k
r)\right).
\end{eqnarray}
The arbitrary constants $C_l$ were determined in such a way that,
far from the origin, the solution is a combination of Hankel
functions of the first kind, i.e., the quantity
$u=\sum_{l=0}^{\infty} u_l P_l(\cos\theta)$ describes
a plane wave $I$ 
\begin{equation}
 I = \sum_{l=0}^{\infty} i^l(2l +1)\sqrt{\frac{\pi}{2kr}}{\cal I}_{l+1/2}(kr)
 P_{l}(\cos\theta),
 \label{41}
\end{equation}
plus a diverging wave $S=u-I$
\begin{equation}
 S = \sum_{l=0}^{\infty} i^l(2l +1)\frac{\epsilon_l}{\sqrt{r}}H^1_{l+1/2}(kr)
 P_{l}(\cos\theta),
 \label{42}
 \end{equation}
 with $H^1_{l+1/2}(kr)={\cal I}_{l+1/2}(kr) +i{\cal
N}_{l+1/2}(kr)$, hence he obtained
\begin{eqnarray}
 C_l &=& \frac{i^l}{a+ib}(2l +1)\sqrt{\frac{\pi}{2k}}\nonumber \\
\epsilon_l&=& -\frac{2ibi^l}{a+ib}\frac{(2l
+1)}{2}\sqrt{\frac{\pi}{2k}} \nonumber \\
&=& (e^{2i\theta_l}-1)i^l\frac{2l+1}{2}\sqrt{\frac{\pi}{2k}}
 \label{4}
\end{eqnarray}
where the angle $\theta_l$ is the relative phase between $u_l$ and
${\cal I}_{l+1/2}$ at large distances, which determines the effect
of the scattering center on the {\em l}-th order
\begin{equation}
\tan{\theta_l} = -b_l/a_l,
\end{equation}
and it gives information about the collision cross-section as we
have seen in previous section.

\section{Conclusion}
As emerged from the above Majorana, as early as 1932, motivated by
the experiments of that period, aimed at describing neutron-proton
scattering developed a model of neutron without spin, forerunner
of the upcoming model of neutron plus
spin \cite{majio}. 
In Sect. 1 we have given a review of the principal experiments on
the neutron collisions, in Sect. 2 we have made a summary on the
collision theory. In Sect. 3 we have outlined the proposal of
models of neutrons. Hence we have analyzed the model of complex
neutron proposed by Massey to describe the passage of neutrons
through matter using theory of collision in presence of a
particular expression of potential. Furthermore we have exhibited
the work of Majorana which considered the elementary nature of the
neutron unlike Massey, obtaining the explicit expression of
solutions of wave equation in a square potential well. We have
stressed his original contribution about the analysis of a model
for the neutron as an elementary particle with respect to the
suggestion of Rutherford that there might exist a neutral particle
formed by the close combination of a proton and an electron,
because it was at first natural to suppose that the neutron might
be such a complex particle. Majorana took into account the
difficulties raised by the experimental results on neutron
structure: is the neutron particle composed of one proton plus one
electron closely related, or an elementary neutral particle? He
tried to establish the interaction law between the proton and
neutron based on criteria of simplicity, chosen in such a way as
to allow you to display as correctly as possible the properties of
more general characteristics of the neutron interactions.

\end{document}